\documentclass[conference]{IEEEtran}
\usepackage{cite}
\usepackage{amsmath,amssymb,amsfonts}
\usepackage{algorithmic}
\usepackage{algorithm}
\usepackage{graphicx}
\usepackage{textcomp}
\usepackage{xcolor}
\renewcommand{\baselinestretch}{1}
\usepackage{booktabs} 
\usepackage{epsfig}
\usepackage[TABBOTCAP]{subfigure}
\usepackage{tabularx}
\usepackage{graphicx}
\usepackage{color}
\usepackage{xspace}
\usepackage{thumbpdf}
\usepackage{listings}
\usepackage{verbatim}
\usepackage{booktabs}
\usepackage{colortbl}
\usepackage{url}
\usepackage{nth}
\usepackage{multirow}
\usepackage{multicol}
\usepackage[english]{babel}
\usepackage[utf8]{inputenc}
\usepackage{algorithm}
\usepackage{mathtools}

\def\BibTeX{{\rm B\kern-.05em{\sc i\kern-.025em b}\kern-.08em
    T\kern-.1667em\lower.7ex\hbox{E}\kern-.125emX}}
\begin{document}
\title{Adversarial ML Attack on Self Organizing Cellular Networks}

\author{\IEEEauthorblockN{Salah-ud-din Farooq \IEEEauthorrefmark{1},
Muhammad Usama\IEEEauthorrefmark{1},
Junaid Qadir\IEEEauthorrefmark{1},
Muhammad Ali Imran\IEEEauthorrefmark{2}
}
\IEEEauthorblockA{\IEEEauthorrefmark{1}Information Technology University, Lahore, Punjab, Pakistan}
\IEEEauthorblockA{\IEEEauthorrefmark{2}University of Glasgow, Scotland, UK}

Email: \IEEEauthorrefmark{1}(mscs16027, muhammad.usama, junaid.qadir)@itu.edu.pk, 
\IEEEauthorrefmark{2}muhammad.imran@glasgow.ac.uk
}

\maketitle

\begin{abstract}
Deep Neural Networks (DNN) have been widely adopted in self-organizing networks (SON) for automating different networking tasks. Recently, it has been shown that DNN lack robustness against adversarial examples where an adversary can fool the DNN model into incorrect classification by introducing a small imperceptible perturbation to the original example. SON is expected to use DNN for multiple fundamental cellular tasks and many DNN-based solutions for performing SON tasks have been proposed in the literature have not been tested against adversarial examples. In this paper, we have tested and explained the robustness of SON against adversarial example and investigated the performance of an important SON use case in the face of adversarial attacks. We have also generated explanations of incorrect classifications by utilizing an explainable artificial intelligence (AI) technique.   

\end{abstract}


\section{Introduction}

Driven by ambitious bandwidth and latency targets, and the development of new domains such as IoT and connected vehicles,  5G networks are becoming increasingly complex as they incorporate disparate emerging trends such as network densification and co-existence with existing cellular technologies. These networks also perform several challenging activities---such as planning, dimensioning, deployment, testing, network optimization, comprehensive performance monitoring, failure detection, failure correction, and general maintenance---which currently utilize large human resources in the loop. This results in a network that is both costly--thus dissatisfying for the cellular operator and error-prone--bringing customer dissatisfaction and resulting in increased churn \cite{ramiro2011self}). 

In such scenarios, artificial intelligence (AI) driven self-organized networks provides an attractive alternative by providing the tools for performing automation with self-organization and intelligence. The main objectives of the SON are to build an intelligent network that can guarantee the network resilience with reduced complexity, simplified network management, and properly optimized network configurations \cite{klaine2017survey}. SON technology leverages advance in machine learning (ML) and deep learning (DL) techniques to overcome the multiple challenges of operating modern network through their integral capability of handling and analyzing big data. 

Even though ML and DL models can greatly outperform traditional methods in obtaining excellent accuracy in benign environments, it is also important to verify the robustness of these models in adversarial settings, particularly since it has been shown in recent work that \textit{adversarial examples} can be generated by malicious adversaries to fool the DL models very easily by applying small perturbations to the original inputs \cite{42503, 43405}. More formally, an adversarial sample $x^*$ is created by following the equation \ref{eq1}, where imperceptible perturbation is denoted as $\delta$; legitimate test example is denoted as $x$; the deployed trained classifier is described by $f(.)$; and $t$ describes the wrong class adversary wants to achieve. 

 \begin{equation}\label{eq1}
     x^* = x + \arg \underset{\eta{_x}}{\text{min}} \{\|\eta\|: f(x + \eta) = t\} 
 \end{equation}

Deep neural networks (DNN) work in a black box manner and this lack of transparency can be a major drawback for the security critical domains. Hence, explainable AI (XAI) or black-box model interpretability plays an important part in mitigating this threat of adversaries. Tomsett et al. \cite{8455710} proposed this phenomenon that XAI and adversarial machine learning (AML) are conceptually linked, and insights into one ot them can provide insights into the other. This is because most vulnerable features after adversarial attacks, together with the help of XAI and AML, can be identified and ultimately any relevant defensive technique can be applied. This interpretability becomes more important now because of the recent adaptation of explainable AI at government levels like General Data Protection Regulation (GDPR)\footnote{GDPR is an EU law regulation aiming at data protection and privacy for all individual citizens of the European Union and the European Economic Area.}, which expresses the importance of explanations of the logic involved when automated decision making takes place \cite{guidotti2018survey}.

The main contributions of our work are; 
\begin{itemize}
    \item Experimentally validated the impact of adversarial attacks in the domain of SON.
    \item Demonstrated that the explainable AI and adversarial ML are linked with each other and adversarial ML can be used to describe feature representations of a DNN model. 
    \item To the best of our knowledge, this study is first in the domain of SON to test adversarial machine learning (AML).  
\end{itemize}




    
    
    

In the section II, we have provided a brief review of the related research that focuses on SON, adversarial ML, and explainable AI. Section III describes the methodology where we have discussed the assumed threat model, ML-models used for a SON use case of detection of abnormal key performance indicator (KPI), and dataset details used in this experiment.  Section IV provides the performance evaluation of the adversarial attacks on the abnormal KPI detector before and after the adversarial attack. Section IV provides the results of adversarial training used as a defense against adversarial attacks. Section V concludes the study and provides future directions.

\section{Related Work}
\label{rw} 
\subsection{SON}
To provide best cellular services to the end-users, efficient network optimization is a continuous process of planning, parametric configuration changes, operations and maintenance with the help of large human interventions. Therefore, the SON is introduced as an intelligent network that provides scalability, agility, and stability to maintain the operators' and consumers' desired objectives \cite{aliu2013survey}. A fundamental property of the SON is the ability to interact and learn from the networking environment to adapt to the changing circumstances. 

Three main functions of SON: self-configuration, self-healing, and self-optimization perform these automatic tasks \cite{3GPP:2011}. 
Self-configuration manages tasks of automatic configuration of cellular network nodes. The main use cases are planning and modifying the radio and transport parameters. Self-optimization manages solutions that target cellular network performance optimization based on the operator specifications. The main use cases of this category are handover parameters optimization, QoS-related parameters optimization, and load balancing. Whereas self-healing manages tasks to automatic detection and rectification of failures in network. 

In the context of cellular systems, DNNs are applied in all three categories of SON. Feng et al. \cite{7446842} used DNN to implement Cell Outage Detection. Daroczy et al. \cite{7145925} used ML to predict Radio Access Bearer (RAB) sessions drops well before the end of the session. Other important work for SON in cellular networks using DNN include resource optimization \cite{4562719}, and mobility management \cite{Majumdar:2005:MUT:1160086.1160094}. Recently, Chen et al. \cite{chen2019unsupervised} combined adversarial training with variational autoencoders to unsupervised learning the behavior of abnormal KPI on the Internet. 

An adversary can affect DNN models of SON through internal and external attacks. In the case of internal attacks, adversaries can corrupt training data and classifiers of DNN models of SON directly. However, these internal attacks are not easily possible due to the difficult task of adding adversarial examples directly into the input of the DNN model. Whereas, external attacks can utilize vulnerabilities of data collection process of cellular networks. Base stations collect measurement reports and pass it SON function that uses this collected data to implement its different functionalities for network optimization. Adversarial examples can be injected into this data collection process with the help of a rogue base station. Shaik et al \cite{Shaik:2018:IRB:3212480.3212497} demonstrated the security vulnerabilities of SON-enabled LTE networks. They injected the fake data into the SON ecosystem with the help of a rogue base station. There work is mainly concerning DoS attacks on cellular networks and user devices. 

\subsection{Adversarial Attacks on SONs and Cognitive Networks}
Most of the current research of adversarial machine learning is relevant to computer vision tasks, such as Szegedy et al. \cite{42503} shows that deep neural network can change its prediction by using non-random perturbation in its inputs. These changes are imperceptible due to the extremely low probability of negative adversaries in every test set. Goodfellow et al. \cite{43405} and Papernot et al. \cite{papernot2017practical} extended this initial study and proposed Fast Gradient Sign Method (FGSM) and Jacobian-based Saliency Map Attack (JSMA) respectively for generating adversarial examples. FGSM is a technique for crafting an adversarial example where one step gradient update is performed in the direction of the sign associated with the gradient at each feature in the test example. The FGSM perturbation ($\eta$) is given as: 

\begin{equation} \label{eq2}
\eta =\epsilon \textit{sign}(\nabla_x j_\theta(x,l))
\end{equation}

Whereas JSMA is based on the concept of saliency maps. This algorithm tries to find input dimensions or features that are most vulnerable due to possible perturbations by creating a saliency map and an iterated process to find misclassification in the model. 

\begin{equation} \label{eq5}
J(x) = \frac{\partial f(x)}{\partial x} = [\frac{\partial f{_j}(x)}{\partial (x{_i})}] 
\end{equation}

Some recent studies of adversarial examples are performed in the field of network intrusion detection systems (NIDS). In these studies, significant degradation in accuracy is observed for intrusion detection systems after exposing DNNs to adversarial examples \cite{wang2018deep}. Whereas, Usama et al. \cite{usama2018adversarial} investigated the vulnerability of Cognitive Self Organizing Networks (CSON) utilizing ML/DL techniques against adversarial attacks.  In this study, we have performed FGSM and JSMA attack on DNN-based abnormal KPI detector to show that adversarial attacks can be fatal for this important use case of SON.  

\subsection{Adversarial Defense Methods}
Many methods have been proposed for making ML models more robust and mitigating adversarial examples. Adversarial Training \cite{43405} and Defensive Distillation \cite{DBLP:journals/corr/PapernotMWJS15} are two famous defense techniques. We have implemented Adversarial Training as a defensive technique for our experiments. The basic idea of Adversarial Training is to train the model using adversarial examples and assign the same labels of the original examples to the adversarial examples.  

\subsection{Explainable AI}
Current work of explainable AI or black-box model interpretability lies within two categories: global and local interpretability. Global interpretability describes the understanding of the whole logic of a model and follows the entire reasoning leading to all the different possible outcomes. Whereas local interpretability is used to generate an individual explanation to justify why the model made a specific decision for an instance \cite{adadi2018peeking}. Some recent studies explored the link between XAI and AML. Tomsett et al. \cite{8455710} proposed this phenomenon that XAI and adversarial machine learning (AML) are conceptually linked, and insights in one can provide insights in the other domain. Giurgiu et al. \cite{giurgiu2019explainable} used recurrent neural networks and attention mechanism for explaining the failure predictions in time series data. Marino et al. \cite{8591457} proposed a methodology to explain incorrect classifications made by Intrusion Detection Systems (IDS) using the adversarial approach. In this paper, we have used explainable AI to provide a deeper understanding of the features involved in the adversarial ML attack on DNN-based KPI detector.

\section{Methodology}

In this section, we will describe our procedure for performing two types of adversarial attacks on abnormal KPI detector but before that, we will describe the threat model and the dataset used in this experiment.  

\subsection{Threat Model}
This subsection describes the major assumptions considered for performing an adversarial attack on the use case of SON.
\begin{enumerate}
    \item \textit{Adversary Knowledge:} We have used two white-box attack algorithms, which mean adversary has complete knowledge about the model architecture, features, and test data. 
    \item \textit{Adversarial Goals and Defense:} Our goal in this experiment is to check the vulnerabilities of SON against adversarial examples. We have achieved this by measuring accuracy before and after implementing attacks. We have experimentally validated a defensive technique to mitigate the effect of adversarial examples. 
\end{enumerate}

\subsection{SON Use Case - Detection of Abnormal KPI}
Figure. 1 states the generic flow chart of SON methodology consisting of its main use cases of self-optimization and Self-configuration of a LTE (Long Term Evolution) network \cite{Shaik:2018:IRB:3212480.3212497}. Two main functions of LTE architecture are (i) Evolved Universal Terrestrial Radio Access Network (E-UTRAN) and (ii) Evolved Packet Core (EPC). E-UTRAN consists of multiple base stations, termed as e-NodeB and User Equipment (UE). UE is typically a smartphone or an IoT device for using call or data services after setting up a connection to a cell of the cellular network. A Cell is a specific terrestrial area controlled by each eNodeB.  

Key Performance Indicators (KPIs) explain the quality of services (QoS) and quality of experience (QoE) of these connected devices. For example, KPIs that are relevant to call or data services setup and services completion belong to Accessibility and Retainability classes of KPIs respectively. The calculation of these KPIs is based on the measurement reports, which are collected through various internal and external data collection methods. SON continuously monitors these KPIs and in case of any abnormality, automatically starts relevant optimization and configuration tasks. 

\begin{figure}[h]
\centering     
\centerline{\includegraphics[width=0.44\textwidth]{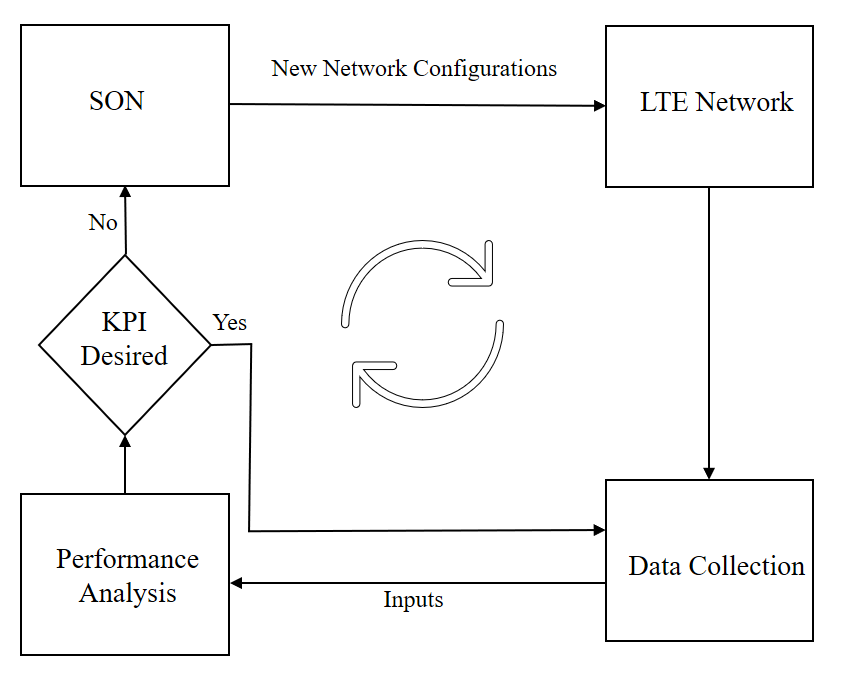}}
\caption{SON Methodology Flow Chart. SON function continuously monitors the KPIs and starts automatic optimization and configuration actions based on KPI measurements}
\end{figure}

E-RAB Drop Rate is one of the significant KPIs to judge user experience and belongs to the retainability class of KPIs. Effective and timely detection of this indicator is essential to avoid users churn. Brief description of this KPI is mentioned below: 

When a User Equipment (UE) has data to send or receive, it sets up an end to end communication channel called EPS Radio Access Bearer (E-RAB) between itself and the core network part of EPC. This E-RAB is the access layer bearer for carrying service data of a UE. After the utilization of cellular services, UE releases its Radio Access Bearer (RAB). This RAB is considered as a \textit{drop} if it is released abnormally i.e. ongoing session is dropped requiring the user to initiate a new connection to resume services. This drop rate is measured as a fraction of the total number of abnormal releases with normal releases.

\[ ERAB\,Drop\,Rate\; \% = 100 \times \frac{{ERAB\,Abnormal\, Releases}}{ERAB\,Normal\,Releases}\] \
    
\subsection{Dataset and Data Pre-processing}

For the use case of E-RAB Drop Rate detection, records are extracted from live LTE network. Each row contains an hourly record of a specific eNodeB with a sudden increase in E-RAB Drop Rate is labeled as an anomaly. Initial experiments involve total 4464 records of two LTE eNodeBs. 3940 records are labeled as normal and 524 as anomalies based on domain knowledge. Each sample has 22 features, which are divided into three main categories of (i) time and location, (ii) dependent features (E-RAB drop reasons) and (iii) independent features (signal strengths, latency, and the number of users). This dataset has binary and nominal data variables and we have applied one-hot encoding to convert nominal features to numeric features since DNNs cannot operate on nominal data directly. This resulted in the transformation of the 25-feature dataset into a 26-feature dataset after one-hot encoding.

After analyzing the data, we have noticed varying distributions of each feature. For example, the mean and standard distribution of some features are larger by seven orders of magnitude from some other features. Without performing normalization, these features would dominate other features. To mitigate this effect, we have used min-max scaling using Scikit-learn library to normalize data. For our use case of anomaly detection in the dataset of E-RAB Drop Rate, we have used Multilayer Perceptron (MLP) classifier with the activation function of ReLU using Keras and Tensorflow sequential model. The MLP model is composed of three hidden layers of 256 neural units. The output layer contains two neurons since labels have two normal and abnormal classes. For regularization, dropout with a rate of 0.4 and early-stopping is used.

\section{Performance Evaluation}
\label{ee}
In this section, we have provided a detailed evaluation of our experiment results. 

\subsection{Evaluation Metric}
We have used accuracy for performance evaluation of results. Accuracy is defined as the percentage of correctly classified records over the total number of records. After training the DNN model, and testing its accuracy, we have implemented both FGSM and JSMA attacks for evaluation of the impact of adversarial examples at the dataset. Accuracy is again measured after implementation of Adversarial Training as a defensive technique. 

\subsection{Experiment Results}
\subsubsection{Impact on Accuracy}
Figure 1 and figure 2 describe the experimental results after implementing adversarial examples and defensive technique of adversarial training at the dataset. It is clear that adversarial examples have significantly degraded the performance of DNNs used in SON. We have observed JSMA caused more performance degradation than FGSM. However, JSMA requires more computation time for crafting an adversarial example. Our results in figure 1 and 2 also depict the performance of DNN-based abnormal KPI after the adversarial training. It is evident from the results that adversarial training has performed better against FGSM as compared to JSMA. 

\begin{figure}[h]
\centering     
\centerline{\includegraphics[width=0.48\textwidth]{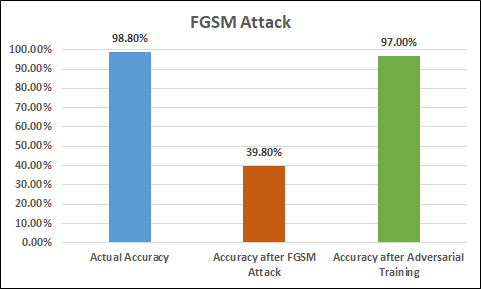}}
\caption{Performance of abnormal KPI detector before and after FGSM attack. The figure also provides the results of adversarial training which tells the recovery of the abnormal KPI detector.}
\end{figure}

\begin{figure}[h]
\centering     
\centerline{\includegraphics[width=0.48\textwidth]{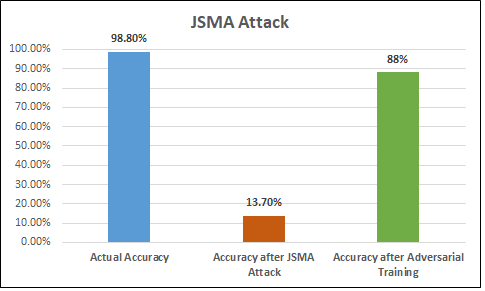}}
\caption{Performance of abnormal KPI detector before and after JSMA attack. The figure also provides the results of adversarial training which tells the recovery of the abnormal KPI detector.}
\end{figure}

\subsubsection{Features Explanations}
Most affected features are calculated through the technique mentioned in \cite{8591457}. We have ranked and sorted the features with their importance after generating the adversarial test set. This importance is calculated by subtracting the original test set from the adversarial test set. The indexes where adversaries have no impact, the value of this subtraction is zero. However, for indexes which are affected by the attack, the value of this subtraction is non-zero. By calculating these non-zero values, most affected features are calculated.

As expected, FGSM changed almost all features (22 out of 25). However, it is not possible to avoid human observation when the scale of the adversary is on such a large level. Whereas JSMA changed six features and degraded the performance of model badly. We have observed the most affected features and compared them with the domain knowledge of cellular networks.  For instance, we have an understanding from the dataset that most of the E-RAB drops are due to the transport network layer (TNL) problems and almost identical features of TNL drops are observed after examining the most vulnerable features by JSMA attack.

\section{Conclusions}
In this paper, we have performed FGSM and JSMA attack on DNN-based abnormal KPI detector. Our results indicate more than 60\% drop in the performance of DNN-based abnormal KPI detector making it very evident that DNN used for detection does not provide robustness against adversarial perturbation. A prominent recovery in the performance of abnormal KPI detector is noticed after we have used adversarial training as a defense against adversarial examples. We have also provided the reasons why adversarial attacks are so effective against abnormal KPI detector by utilizing insights from explainable AI. Our results also trhow the light on a previously ignored area of machine learning security in the SON and provide good insights for developing a robust ML-based SON solution.  
   
\def\bibfont{\small}
\bibliographystyle{IEEEtran}
\bibliography{IEEEabrv,reference}

\begin{thebibliography}{10}
\providecommand{\url}[1]{#1}
\csname url@samestyle\endcsname
\providecommand{\newblock}{\relax}
\providecommand{\bibinfo}[2]{#2}
\providecommand{\BIBentrySTDinterwordspacing}{\spaceskip=0pt\relax}
\providecommand{\BIBentryALTinterwordstretchfactor}{4}
\providecommand{\BIBentryALTinterwordspacing}{\spaceskip=\fontdimen2\font plus
\BIBentryALTinterwordstretchfactor\fontdimen3\font minus
  \fontdimen4\font\relax}
\providecommand{\BIBforeignlanguage}[2]{{%
\expandafter\ifx\csname l@#1\endcsname\relax
\typeout{** WARNING: IEEEtran.bst: No hyphenation pattern has been}%
\typeout{** loaded for the language `#1'. Using the pattern for}%
\typeout{** the default language instead.}%
\else
\language=\csname l@#1\endcsname
\fi
#2}}
\providecommand{\BIBdecl}{\relax}
\BIBdecl

\bibitem{ramiro2011self}
J.~Ramiro and K.~Hamied, \emph{Self-organizing networks: self-planning,
  self-optimization and self-healing for GSM, UMTS and LTE}.\hskip 1em plus
  0.5em minus 0.4em\relax John Wiley \& Sons, 2011.

\bibitem{klaine2017survey}
P.~V. Klaine, M.~A. Imran, O.~Onireti, and R.~D. Souza, ``A survey of machine
  learning techniques applied to self-organizing cellular networks,''
  \emph{IEEE Communications Surveys \& Tutorials}, vol.~19, no.~4, pp.
  2392--2431, 2017.

\bibitem{42503}
\BIBentryALTinterwordspacing
C.~Szegedy, W.~Zaremba, I.~Sutskever, J.~Bruna, D.~Erhan, I.~Goodfellow, and
  R.~Fergus, ``Intriguing properties of neural networks,'' in
  \emph{International Conference on Learning Representations}, 2014. [Online].
  Available: \url{http://arxiv.org/abs/1312.6199}
\BIBentrySTDinterwordspacing

\bibitem{43405}
\BIBentryALTinterwordspacing
I.~Goodfellow, J.~Shlens, and C.~Szegedy, ``Explaining and harnessing
  adversarial examples,'' in \emph{International Conference on Learning
  Representations}, 2015. [Online]. Available:
  \url{http://arxiv.org/abs/1412.6572}
\BIBentrySTDinterwordspacing

\bibitem{8455710}
R.~{Tomsett}, A.~{Widdicombe}, T.~{Xing}, S.~{Chakraborty}, S.~{Julier},
  P.~{Gurram}, R.~{Rao}, and M.~{Srivastava}, ``Why the failure? how
  adversarial examples can provide insights for interpretable machine
  learning,'' in \emph{2018 21st International Conference on Information Fusion
  (FUSION)}, July 2018, pp. 838--845.

\bibitem{guidotti2018survey}
R.~Guidotti, A.~Monreale, S.~Ruggieri, F.~Turini, F.~Giannotti, and
  D.~Pedreschi, ``A survey of methods for explaining black box models,''
  \emph{ACM Computing Surveys (CSUR)}, vol.~51, no.~5, p.~93, 2018.

\bibitem{aliu2013survey}
O.~G. Aliu, A.~Imran, M.~A. Imran, and B.~Evans, ``A survey of self
  organisation in future cellular networks,'' \emph{IEEE Communications Surveys
  \& Tutorials}, vol.~15, no.~1, pp. 336--361, 2013.

\bibitem{3GPP:2011}
3GPP, ``Evolved universal terrestrial radio access network (e-utran);
  self-configuring and self-optimizing network (son) use cases and solutions.
  ts 36.902. 3rd generation partnership project (3gpp).''
  \url{http://www.3gpp.org/DynaReport/36902.htm}, 2011.

\bibitem{7446842}
{Wanrong Feng}, {Yinglei Teng}, {Yi Man}, and {Mei Song}, ``Cell outage
  detection based on improved bp neural network in lte system,'' in \emph{11th
  International Conference on Wireless Communications, Networking and Mobile
  Computing (WiCOM 2015)}, Sep. 2015, pp. 1--5.

\bibitem{7145925}
B.~{Daroczy}, P.~{Vaderna}, and A.~{Benczur}, ``Machine learning based session
  drop prediction in lte networks and its son aspects,'' in \emph{2015 IEEE
  81st Vehicular Technology Conference (VTC Spring)}, May 2015, pp. 1--5.

\bibitem{4562719}
P.~{Sandhir} and K.~{Mitchell}, ``A neural network demand prediction scheme for
  resource allocation in cellular wireless systems,'' in \emph{2008 IEEE Region
  5 Conference}, April 2008, pp. 1--6.

\bibitem{Majumdar:2005:MUT:1160086.1160094}
\BIBentryALTinterwordspacing
K.~Majumdar and N.~Das, ``Mobile user tracking using a hybrid neural network,''
  \emph{Wirel. Netw.}, vol.~11, no.~3, pp. 275--284, May 2005. [Online].
  Available: \url{http://dx.doi.org/10.1007/s11276-005-6611-x}
\BIBentrySTDinterwordspacing

\bibitem{chen2019unsupervised}
W.~Chen, H.~Xu, Z.~Li, D.~Peiy, J.~Chen, H.~Qiao, Y.~Feng, and Z.~Wang,
  ``Unsupervised anomaly detection for intricate kpis via adversarial training
  of vae,'' in \emph{IEEE INFOCOM 2019-IEEE Conference on Computer
  Communications}.\hskip 1em plus 0.5em minus 0.4em\relax IEEE, 2019, pp.
  1891--1899.

\bibitem{Shaik:2018:IRB:3212480.3212497}
\BIBentryALTinterwordspacing
A.~Shaik, R.~Borgaonkar, S.~Park, and J.-P. Seifert, ``On the impact of rogue
  base stations in 4g/lte self organizing networks,'' in \emph{Proceedings of
  the 11th ACM Conference on Security \& Privacy in Wireless and Mobile
  Networks}, ser. WiSec '18.\hskip 1em plus 0.5em minus 0.4em\relax New York,
  NY, USA: ACM, 2018, pp. 75--86. [Online]. Available:
  \url{http://doi.acm.org/10.1145/3212480.3212497}
\BIBentrySTDinterwordspacing

\bibitem{papernot2017practical}
N.~Papernot, P.~McDaniel, I.~Goodfellow, S.~Jha, Z.~B. Celik, and A.~Swami,
  ``Practical black-box attacks against machine learning,'' in
  \emph{Proceedings of the 2017 ACM on Asia Conference on Computer and
  Communications Security}.\hskip 1em plus 0.5em minus 0.4em\relax ACM, 2017,
  pp. 506--519.

\bibitem{wang2018deep}
Z.~Wang, ``Deep learning-based intrusion detection with adversaries,''
  \emph{IEEE Access}, vol.~6, pp. 38\,367--38\,384, 2018.

\bibitem{usama2018adversarial}
M.~Usama, J.~Qadir, and A.~Al-Fuqaha, ``Adversarial attacks on cognitive
  self-organizing networks: The challenge and the way forward,'' in \emph{2018
  IEEE 43rd Conference on Local Computer Networks Workshops (LCN
  Workshops)}.\hskip 1em plus 0.5em minus 0.4em\relax IEEE, 2018, pp. 90--97.

\bibitem{DBLP:journals/corr/PapernotMWJS15}
\BIBentryALTinterwordspacing
N.~Papernot, P.~D. McDaniel, X.~Wu, S.~Jha, and A.~Swami, ``Distillation as a
  defense to adversarial perturbations against deep neural networks,''
  \emph{CoRR}, vol. abs/1511.04508, 2015. [Online]. Available:
  \url{http://arxiv.org/abs/1511.04508}
\BIBentrySTDinterwordspacing

\bibitem{adadi2018peeking}
A.~Adadi and M.~Berrada, ``Peeking inside the black-box: A survey on
  explainable artificial intelligence (xai),'' \emph{IEEE Access}, vol.~6, pp.
  52\,138--52\,160, 2018.

\bibitem{giurgiu2019explainable}
I.~Giurgiu and A.~Schumann, ``Explainable failure predictions with rnn
  classifiers based on time series data,'' \emph{arXiv preprint
  arXiv:1901.08554}, 2019.

\bibitem{8591457}
D.~L. {Marino}, C.~S. {Wickramasinghe}, and M.~{Manic}, ``An adversarial
  approach for explainable ai in intrusion detection systems,'' in \emph{IECON
  2018 - 44th Annual Conference of the IEEE Industrial Electronics Society},
  Oct 2018, pp. 3237--3243.

\end{thebibliography}

\end{document}